\documentclass{llncs}
\usepackage{paralist}
\usepackage{graphicx}
\usepackage{url}

\usepackage{xcolor}
\usepackage[colorlinks,citecolor=blue,urlcolor=gray]{hyperref}

\title{The MMT API: A Generic MKM System}
\author{Florian Rabe}
\institute{Computer Science, Jacobs University Bremen, Germany\\ 
\protect\url{http://trac.kwarc.info/MMT}
}

\newcommand{\mmt}{{\sc Mmt}}
\newcommand{\openmath}{{\sc OpenMath}}
\newcommand{\omdoc}{{\sc OMDoc}}
\newcommand{\tntbase}{{\sc TNTBase}}

\begin{document}
\maketitle

\begin{abstract}
The {\mmt} language has been developed as a scalable representation and interchange language for formal mathematical knowledge.
It permits natural representations of the syntax and semantics of virtually all declarative languages while making {\mmt}-based MKM services easy to implement.
It is foundationally unconstrained and can be instantiated with specific formal languages. 

The {\mmt} API implements the {\mmt} language along with multiple backends for persistent storage and frontends for machine and user access.
Moreover, it implements a wide variety of {\mmt}-based knowledge management services.
The API and all services are generic and can be applied to any language represented in {\mmt}. A plugin interface permits injecting syntactic and semantic idiosyncrasies of individual formal languages.
\end{abstract}

\paragraph{The {\mmt} Language}
Content-oriented representation languages for mathematical knowledge are usually designed to focus on either of two goals:
\begin{inparaenum}[(i)]
 \item the automation potential offered by mechanically verifiable representations, as pursued in semi-automated proof assistants like Isabelle and
 \item the universal applicability offered by a generic meta-language, as pursued in XML-based content markup languages like {\omdoc}.
\end{inparaenum}
The {\mmt} language \cite{RK:mmt:10} (Module system for Mathematical Theories) was designed to realize both goals in one coherent system.
It uses a minimal number of primitives with a precise semantics chosen to permit natural and adequate representations of many individual languages.

A key feature is \emph{foundation-independence}: {\mmt} systematically avoids a commitment to a particular type theory or logic.
Instead, it represents every formal system as an {\mmt} theory: domain theories (like the theory $\mathtt{Group}$), logics (like first-order logic FOL), and logical frameworks (like LF \cite{lf}) are represented uniformly as {\mmt} \emph{theories}.
These theories are related by the \emph{meta-theory} relation, e.g., LF is the meta-theory of FOL, which in turn is the meta-theory of $\mathtt{Group}$.
{\mmt} uses this relation to obtain the semantics of a theory from that of its meta-theory; thus, an external semantics (called the \emph{foundation}), e.g., a research article or an implementation, only has to be supplied for the topmost meta-theories.
For example, a foundation for LF can be given in the form of a type system.

Theories contain typed \emph{symbol declarations}, which permit the uniform representation of constants, functions, and predicates as well as -- via the Curry-Howard correspondence -- judgments, inference rules, axioms, and theorems.
Theories are related via \emph{theory morphisms}, which subsume translations, functors, and models.
Finally, {\mmt} provides a module system for building large theories and morphisms via reuse and inheritance.

Mathematical objects such as terms, types, formulas, and proofs are represented uniformly as {\openmath} objects \cite{openmath}, which are formed from the symbols available to the theory under consideration: For example, LF declares the symbols $\mathtt{type}$ and $\lambda$; FOL declares $\forall$ and $\Rightarrow$, and $\mathtt{Group}$ declares $\circ$ and $e$.
{\mmt} is agnostic in the typing relation between these objects and instead delegates the resolution of typing judgments to the foundation.
Then all {\mmt} results are obtained for arbitrary foundations.
For example, {\mmt} guarantees that theory morphisms translate objects in a typing- and truth-preserving way, which is the crucial invariant permitting the reuse of results in large networks of theories.

\paragraph{The {\mmt} API}
Exploiting the small number of primitives in {\mmt}, the {\mmt} API provides a comprehensive, scalable implementation of {\mmt} itself and of {\mmt}-based knowledge management (KM) services.
The development is intentionally \emph{application-independent}: It focuses on the data model of {\mmt} and its KM services in a way that makes the integration into specific applications as easy as possible.
But by itself, it provides only a basic user interface.

All algorithms are implemented generically and relegate all foundation-specific aspects to plugins.
Concrete applications usually provide a few small plugins to customize the behavior to one specific foundation and a high-level component that connects the desired {\mmt} services to a user interface.

The API is written in the functional and object-oriented language Scala \cite{scala}, which is fully compatible with Java so that API plugins and applications can be written in either language.
%The API depends on only one external library -- the lean web server tiscaf \cite{tiscaf}.
Excluding plugins and libraries, it comprises over $20,000$ lines of Scala code compiling into about $3000$ Java \texttt{class} files totaling about $5$ MB of platform-independent bytecode.
Sources, binaries, API documentation, and user manual are available at \url{http://trac.kwarc.info/MMT}.

\paragraph{Knowledge Management Services}
The {\mmt} API provides a suite of coherently integrated KM services, which we only summarize here because they have been presented individually.
A \emph{notation language} based on \cite{KMR:presentation:08} is used to serialize {\mmt} in arbitrary output formats. Notations are grouped into \emph{styles}, and a rendering engine presents any {\mmt} concept according to the chosen style.

{\mmt} content can be organized in \emph{archives} \cite{HIJKR:dimensions:11}, a light-weight project abstraction to integrate source files, content markup, narrative structure, notation indices, and RDF-style relational indices. Archives can be built, indexed, and browsed, and simplify distribution and reuse.
A \emph{query language} \cite{rabe:querying:12} integrates hierarchic, relational, and unification-based query paradigms.
A \emph{change management} infrastructure \cite{IR:moc:12} permits detecting and propagating and changes at the level of individual {\openmath} objects.

\paragraph{User and System Interfaces}
If run as a standalone application, the API responds with a shell that interacts via standard input/output.
The shell is scriptable, which permits users and application developers to initialize and configure it conveniently.
For example, to check the theory $\mathit{Group}$, the initialization script would first register the {\mmt} theory defining the syntax of LF and then a plugin providing a foundation for LF, then register the theory FOL, and finally check the file containing the theory $\mathit{Group}$.

A second frontend is given by an HTTP server. For machine interaction, it exposes all API functionality and KM services via a RESTful interface, which permits developing {\mmt}-based applications outside the Java/Scala world.
For human interaction, the HTTP server offers an interactive \emph{web browser} based on HTML+presentation MathML. The latter is computed on demand according to the style interactively selected by the user.
Based on the JOBAD JavaScript library \cite{GLR:jobad:09}, user interaction is handled via JavaScript and Ajax.
In particular, {\mmt} includes a JOBAD module that provides interactive functionality such as definition lookup and type inference.
\medskip

To facilitate distributing {\mmt} content, all {\mmt} declarations are referenced by canonical \emph{logical identifiers} (the {\mmt} URIs), and their physical locations (their URLs) remain transparent. This is implemented as a catalog that translates {\mmt} URIs into URLs according to the registered knowledge repositories.
{\mmt} declarations are retrieved and loaded into memory transparently when needed so that storage and memory management are hidden from high-level services, applications, and users.
Supported knowledge repositories are file systems, SVN working copies and repositories, and {\tntbase} databases \cite{tntbase}.
The latter also supports {\mmt}-specific indexing and querying functions \cite{KRZ:mmttnt:10} permitting, e.g., the efficient retrieval of the dependency closure of an {\mmt} knowledge item.

%For example, the API acts as a {\tntbase} plugin to create {\mmt}-specific indices, which are used by an XQuery module to provide efficient {\mmt}-specific access methods. For example, 

\paragraph{A Specific Application for a Specific Foundation}
The LATIN project \cite{CHKMR:latinabs:11} built an atlas of logics and related formal systems.
The atlas is realized as an {\mmt} project, and {\mmt} is used for building and interactively browsing the atlas.

All theories in the atlas use LF as their meta-theory, which defines the abstract syntax of LF and thus of the logics in the atlas.

For concrete syntax, the Twelf implementation of LF is used.
To integrate Twelf with {\mmt}, LATIN developed an {\mmt} plugin that calls Twelf to read individual source files and convert them to {\omdoc}, which {\mmt} reads natively.

Based on this import, {\mmt}'s foundation-independent algorithms can index and catalog the LATIN atlas and make it accessible to KM services and applications.
Here the use of Twelf remains fully transparent: An application sends only an {\mmt} URI (e.g., the one LATIN defines for the theory FOL) to {\mmt} and receives the corresponding Scala object.

From the perspective of {\mmt}, Twelf is an external tool for parsing and type reconstruction that is applicable only to theories whose meta-theory is LF.
From the perspective of Twelf on the other hand, the {\mmt} theory LF does not exist.
Instead, the symbols $\mathit{type}$, $\lambda$, etc. are implemented directly in Twelf's underlying programming language.

This is a typical situation: Generally, {\mmt} uses the meta-theory to determine which plugin is applicable, and these plugins hard-code the semantics of the respective meta-theory.
Similar concrete syntax plugins can be written for most languages and exist for, e.g., the ATP interface language TPTP, the ontology language OWL, and the Mizar language for formalized mathematics.
\medskip

\begin{figure}[ht]
\includegraphics[width=\textwidth]{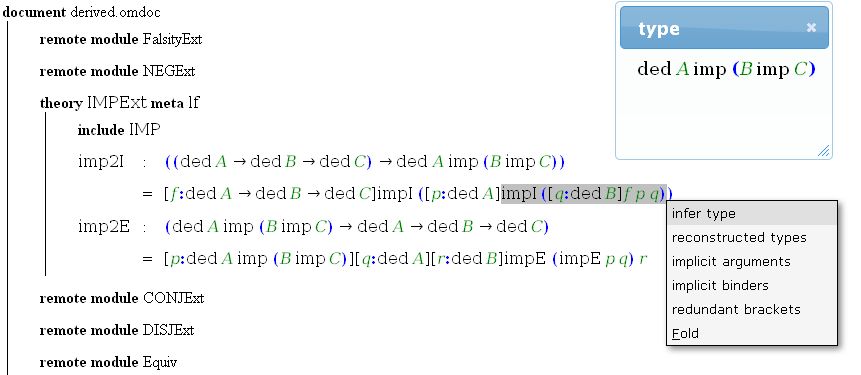}
\end{figure}

LATIN also customizes the {\mmt} web server by providing a style that provides notations for objects from theories with meta-theory LF.
The above screen shot shows the generic web server displaying a theory $\mathtt{IMPExt}$: It imports the theory $\mathrm{IMP}$ of the implication connective $\mathtt{imp}$ and extends it with derived rules for the introduction and elimination of double implications, i.e., formulas of the form $A\, \mathtt{imp}\, (B\,\mathtt{imp}\, \mathtt{C})$.
The symbol $\mathtt{impI}$ represents the derivation of the rule $\frac{A,B\vdash C}{\vdash A\, imp\, (B\,imp\, C)}$.
Via the context menu, the user has called type inference on the selected subobject, which opened a dialog showing the \emph{dynamically inferred type}.

The interactive type inference is implemented using the HTTP interface to the {\mmt} API.
First of all, the LATIN style is such that the rendered HTML includes parallel markup in the form of special attributes on the presentation MathML elements.
JavaScript uses them to build an {\mmt} query that is posted to the server as an Ajax request and whose response is shown in the dialog.
This query bundles multiple {\mmt} API calls into a single HTTP request-response cycle:
First the parallel markup is used to retrieve the {\openmath} object corresponding to the selected expression (and its context), then type inference is called, and finally the rendering engine is called to render the type as presentation MathML.

For the type inference, LATIN provides one further plugin: a foundation plugin that supplies the typing relation for theories with meta-theory LF. {\mmt} uses it to perform type inference directly in memory without having to call external tools like Twelf.
Such foundation plugins are easy to write because they can focus on the logical core of the type system and, e.g., parsing and module system remain transparent to the plugin.
For example, the plugin for LF comprises only 200 lines of code
\medskip

Except for the concrete syntax plugin, the presentation style, and the foundation plugin, all steps of the above example are foundation-independent and are immediately available for {\mmt} content written in any other meta-theory.
Moreover, being a logical framework, these plugins LF are immediately inherited by all logics defined in LF: We obtain, e.g., type inference for FOL and $\mathit{Group}$ (in fact: for all logics defined in LATIN) without writing additional plugins.

Furthermore, all implementations are application-independent and can be immediately integrated into any application, e.g., a Wiki containing LF objects.
This customization of {\mmt} to specific foundations and specific applications occurs at minimal cost, a principle we call \emph{rapid prototyping for formal ystems}.

\paragraph{Acknowledgements}
Over the last 6 years, contributions to the API or to individual plugins have been made by Maria Alecu, Alin Iacob, Catalin David, Stefania Dumbrava, Dimitar Misev, Fulya Horozal, F\"usun Horozal, Mihnea Iancu, Felix Mance, and Vladimir Zamdzhiev. Some of them were partially supported by DFG grant KO-2428/9-1.

\bibliographystyle{plain}
%\bibliography{bib/pub_rabe,bib/rabe,bib/systems}

\begin{thebibliography}{10}

\bibitem{openmath}
S.~Buswell, O.~Caprotti, D.~Carlisle, M.~Dewar, M.~Gaetano, and M.~Kohlhase.
\newblock {The Open Math Standard, Version 2.0}.
\newblock Technical report, The Open Math Society, 2004.
\newblock See \url{http://www.openmath.org/standard/om20}.

\bibitem{CHKMR:latinabs:11}
M.~Codescu, F.~Horozal, M.~Kohlhase, T.~Mossakowski, and F.~Rabe.
\newblock {Project Abstract: Logic Atlas and Integrator (LATIN)}.
\newblock In J.~Davenport, W.~Farmer, F.~Rabe, and J.~Urban, editors, {\em
  Intelligent Computer Mathematics}, pages 289--291. Springer, 2011.

\bibitem{GLR:jobad:09}
J.~Gi{\v{c}}eva, C.~Lange, and F.~Rabe.
\newblock {Integrating Web Services into Active Mathematical Documents}.
\newblock In J.~Carette, L.~Dixon, C.~{Sacerdoti Coen}, and S.~Watt, editors,
  {\em Intelligent Computer Mathematics}, pages 279--293. Springer, 2009.

\bibitem{lf}
R.~Harper, F.~Honsell, and G.~Plotkin.
\newblock {A framework for defining logics}.
\newblock {\em {Journal of the Association for Computing Machinery}},
  40(1):143--184, 1993.

\bibitem{HIJKR:dimensions:11}
F.~Horozal, A.~Iacob, C.~Jucovschi, M.~Kohlhase, and F.~Rabe.
\newblock {Combining Source, Content, Presentation, Narration, and Relational
  Representation}.
\newblock In J.~Davenport, W.~Farmer, F.~Rabe, and J.~Urban, editors, {\em
  Intelligent Computer Mathematics}, pages 212--227. Springer, 2011.

\bibitem{IR:moc:12}
M.~Iancu and F.~Rabe.
\newblock {Management of Change in Declarative Languages}.
\newblock In J.~Campbell, J.~Carette, G.~{Dos Reis}, J.~Jeuring, P.~Sojka,
  V.~Sorge, and M.~Wenzel, editors, {\em Intelligent Computer Mathematics},
  pages 325--340. Springer, 2012.

\bibitem{KMR:presentation:08}
M.~Kohlhase, C.~M{\"u}ller, and F.~Rabe.
\newblock {Notations for Living Mathematical Documents}.
\newblock In S.~Autexier, J.~Campbell, J.~Rubio, V.~Sorge, M.~Suzuki, and
  F.~Wiedijk, editors, {\em Mathematical Knowledge Management}, pages 504--519.
  Springer, 2008.

\bibitem{KRZ:mmttnt:10}
M.~Kohlhase, F.~Rabe, and V.~Zholudev.
\newblock {Towards MKM in the Large: Modular Representation and Scalable
  Software Architecture}.
\newblock In S.~Autexier, J.~Calmet, D.~Delahaye, P.~Ion, L.~Rideau, R.~Rioboo,
  and A.~Sexton, editors, {\em Intelligent Computer Mathematics}, pages
  370--384. Springer, 2010.

\bibitem{scala}
M.~Odersky, L.~Spoon, and B.~Venners.
\newblock {\em {Programming in Scala}}.
\newblock artima, 2007.

\bibitem{rabe:querying:12}
F.~Rabe.
\newblock {A Query Language for Formal Mathematical Libraries}.
\newblock In J.~Campbell, J.~Carette, G.~{Dos Reis}, J.~Jeuring, P.~Sojka,
  V.~Sorge, and M.~Wenzel, editors, {\em Intelligent Computer Mathematics},
  pages 142--157. Springer, 2012.

\bibitem{RK:mmt:10}
F.~Rabe and M.~Kohlhase.
\newblock {A Scalable Module System}.
\newblock {\em Information and Computation}, 2013.
\newblock conditionally accepted; see \url{http://arxiv.org/abs/1105.0548}.

\bibitem{tntbase}
V.~Zholudev and M.~Kohlhase.
\newblock {TNTBase: a Versioned Storage for XML}.
\newblock In {\em {Proceedings of Balisage: The Markup Conference 2009}},
  volume~3 of {\em {Balisage Series on Markup Technologies}}. Mulberry
  Technologies, Inc., 2009.

\end{thebibliography}

\end{document}